\documentclass[conference]{IEEEtran}
\IEEEoverridecommandlockouts
\usepackage{cite}
\usepackage{amsmath,amssymb,amsfonts}
\usepackage{algorithmic}
\usepackage{graphicx}
\usepackage{textcomp}
\usepackage{xcolor}
\usepackage{subcaption}

\usepackage{amsthm}
 
\usepackage{algorithm,algorithmic}
\usepackage{booktabs}
\usepackage{arydshln} 

\def\BibTeX{{\rm B\kern-.05em{\sc i\kern-.025em b}\kern-.08em
    T\kern-.1667em\lower.7ex\hbox{E}\kern-.125emX}}

\begin{document}

\title{
Beyond Energy Functions and Numerical Integration: A New Methodology to Determine Transient Stability at the Initial State



}


\author{
\IEEEauthorblockN{Wenhao Wu$^1$, Dan Wu$^{1}$, Bin Wang$^2$, Jiabing Hu$^1$}
\IEEEauthorblockA{
\textit{1. School of Electrical and Electronic Engineering, Huazhong University of Science and Technology, Wuhan, China}\\
\textit{2. Advanced Technology Solutions, ISO New England, Holyoke, MA, U.S.}
} 
\textit{\{m202572375, danwuhust, j.hu\}@hust.edu.cn and bwang@iso-ne.com}
}

\maketitle


\begin{abstract}
This paper presents a novel method for transient stability analysis (TSA) that circumvents the limitations of sequential numerical integration and energy functions. 
The proposed method begins by constructing a trajectory-dependent stability indicator function to distinguish the system's destiny. 
To overcome the difficulty in analyzing the asymptotic behavior at infinite time, a strategic time contraction mapping is then applied. This allows TSA to be recast as a pole-placement detection problem for the indicator function. 
By leveraging high-order derivatives at the initial state, a rational function approximation is derived, yielding a mathematically direct and computationally efficient prediction.
Numerical validations on benchmark systems demonstrate that the method not only provides a direct mathematical shortcut for TSA in power systems but also establishes a promising new methodology for evaluating the transient stability of a broad class of nonlinear dynamical systems.

\end{abstract}


\section{Introduction}
 Transient stability analysis, which seeks to determine whether a power system remains stable after a disturbance, is a fundamental tool for securing system operations~\cite{kundur2007power}. Given the inherent nonlinearity and high dimensionality of dynamical responses in power grids, transient stability analysis usually needs to make a compromise between accuracy and efficiency. Hence, different methodologies have been developed. 

Time-domain simulations are the most commonly used method in practice. 
It applies numerical integration techniques to replicate the complete trajectory of system responses \cite{hairer1996solvingODE}. 
While favored for its generality and high fidelity, most numerical integration techniques must be carried out in a temporally sequential manner, i.e., computing the next time snapshot from the last one. Such a step-by-step nature prevents parallel processing advantages: the total simulation time is the sum of execution times for all the steps. This inherent sequential processing nature is becoming a major bottleneck for fast stability evaluation of today's power grids with a burst of system size and complexity.

Another methodology that has been widely investigated is the energy-function-based method. It was first introduced to power system transient stability analysis in 1947 \cite{magnusson1947transient}, even before the introduction of digital computing in power systems \cite{dunstan1954digital}. Instead of reconstructing the post-disturbance state trajectory, the energy-function-based method determines the system's stability by comparing its transient energy at the initial point to a critical energy threshold. If the initial energy lies below this threshold, stability is guaranteed. 
Although conceptually appealing and mathematically elegant, this approach suffers from a well-known limitation: finding a valid energy function for general nonlinear dynamical systems is extremely challenging. Considerable efforts have been devoted to deriving such functions for specific classes of power system models \cite{caliskan2014compositional}, from which certain attempts have further moved towards a numerical approach instead of analytical derivations \cite{chang1995numericaldirect}. Moreover, for many practical systems with dissipation, it has been shown that no suitable energy function exists \cite{chiang2002study}. 
On the other hand, the critical energy threshold is usually determined from the closest or controlling type-1 unstable equilibrium point (UEP). Finding these UEPs is another challenging problem, which constitutes a research topic in its own right \cite{wu2019holomorphic, dan2019solutions}.
Therefore, significant limitations and challenges must be overcome before it can become a practical tool~\cite{chiang2011direct}. 

Besides the above model-based methodologies, recent developments in data science and machine learning have enabled a new methodological trend of data-driven methods for transient stability analysis \cite{sarajcev2022artificial}. 
While offering the promise of high computational efficiency, these approaches face significant practical hurdles, including the high cost of data acquisition and labeling~\cite{li2022deep}, the loss of interpretability~\cite{guidotti2018survey}, the limited adaptability to new scenarios, and the poor transferability among different systems~\cite{bellizio2022changingtopology}. 
Motivated by the above challenges and limitations, this work seeks to develop a new methodology for transient stability analysis that is numerical integration-free, easy to derive, and physically meaningful.

\section{Methodological Motivations and Rationale}
At the core of our methodology is a fundamental principle of deterministic dynamical systems: if the vector field $f(x,t)$ is continuous in $t$ and globally Lipschitz in $x$, the future evolution of the system is uniquely determined from the initial state. 
This theoretical determinism implies that the initial state itself contains all the information to ascertain the system's ultimate fate. The only question is \textit{how to extract the stability-related information from an initial state.} 

To answer the above question, time-domain simulations adopt the ``brute-force'' approach by reconstructing the entire state trajectory from the initial state. By reconstructing the full trajectory, this methodology ensures generality and reliability, albeit at the cost of computational efficiency. However, if the sole objective is to assess stability, most of the computational effort spent on obtaining additional information is wasted, resulting in a fundamental inefficiency. 

On the other hand, the energy-function-based methodology is driven by the idea of directly determining whether an initial state lies within the region of attraction of the post-disturbance equilibrium. This is achieved by evaluating a suitably defined energy metric whose value on the stability boundary point serves as the threshold for stability. It requires minimal information from any specific trajectory, but an explicit energy function can be constructed only for systems with particular structures, which is generally not feasible.

The proposed method seeks to leverage ideas from both time-domain simulations and energy-function-based approaches. Given that the state trajectory can reveal the ultimate fate of the system, a trajectory-dependent stability indicator may be derived to extract the stability information. This indicator must be general enough to accommodate a broad class of nonlinear dynamical systems, and sufficiently selective to make a clear distinction between stability and instability, as elaborated in Section~\ref{subsect:3-1}. Inspired by the benign global properties of analytic functions at any regular point, we can derive the Taylor series of the indicator function at the initial state to avoid the temporally sequential computations required by numerical integration. However, a limited radius of convergence cannot reveal the asymptotic behavior of the indicator, and a finite truncation of the Taylor series may misbehave even before the time approaches infinity. Hence, our conceptual leap is to compress the infinite time horizon to a finite interval by a time contraction mapping, as detailed in Section~\ref{subsect:3-2}.  
Finally, an appropriate detector of the asymptotic behavior of the indicator function is needed, which is achieved by a rational function approximation and the associated zeros and poles in Section~\ref{subsect:3-3}. 

Thus, the original transient stability problem is recast as detecting the pole-placement behavior of the indicator function as it approaches its finite time horizon. 
This allows us to use local derivatives of the indicator at the initial state to predict the indicator's asymptotic behavior, leading to a direct mathematical shortcut to the system's infinite-time destiny.

\section{Technical Details of the Proposed Transient Stability Methodology}
\subsection{Transient Stability Indicator} \label{subsect:3-1}
The post-disturbance dynamics of a power system are described by a set of nonlinear DAEs of the form:
\begin{equation}
\begin{aligned}
\dot{\mathbf{x}} &= \mathbf{f}(\mathbf{x}, \mathbf{v}) \\
\mathbf{0} &= \mathbf{g}(\mathbf{x}, \mathbf{v})
\label{eq:DAE}
\end{aligned}
\end{equation}
where $\mathbf{x} \in \mathbb{R}^n$ is the vector of state variables, $\mathbf{v} \in \mathbb{R}^m$ is the vector of bus voltages, $\mathbf{f}$ represents the vector field of the system's dynamic components (e.g., generators and controllers), $\mathbf{g}$ is the power flow equations expressed in rectangular coordinates.

Transient stability assessment fundamentally requires determining the asymptotic behavior of the state trajectory $\mathbf{x}(t)$ originating from a post-fault initial condition $\mathbf{x}(0)$. Specifically, the system is deemed stable if its state converges to the desired post-fault stable equilibrium point (SEP), denoted as $\mathbf{x}^*$.
\begin{equation}
\text{Stable} \iff \lim_{t \to +\infty} \mathbf{x}(t) = \mathbf{x}^*
\end{equation}
Conversely, if the trajectory converges to a different equilibrium point or diverges entirely, the system is unstable.

To transform this asymptotic convergence problem into an analytical test, we introduce a scalar function, hereafter referred to as the \textit{indicator function}, $h(t)$:
\begin{equation}
h(t) = \frac{-1}{\|\mathbf{x}(t) - \mathbf{x}^*\|^2}
= \frac{-1}{(\mathbf{x}(t) - \mathbf{x}^*)^{\top}(\mathbf{x}(t) - \mathbf{x}^*)}
\label{eq:h_t}
\end{equation}
where $h: \mathbb{R} \to \mathbb{R}$. The denominator of $h(t)$ has a clear physical interpretation, i.e., the squared Euclidean distance between the system state $\mathbf{x}(t)$ and the target SEP $\mathbf{x}^*$, providing a sharp distinction between stable and unstable outcomes:

\begin{itemize}
    \item Stable: As $\mathbf{x}(t) \to \mathbf{x}^*$, the denominator approaches zero, which yields $h(t) \to -\infty$.  
    \item Unstable:
    \begin{itemize}
        \item If the trajectory converges to a different equilibrium point ($\mathbf{x}(t) \to \mathbf{x}_{\text{other}} \neq \mathbf{x}^*$), the denominator converges to a finite positive constant, and $h(t)$ approaches a finite negative value.
        \item If the trajectory diverges ($\|\mathbf{x}(t)\| \to \infty$), the denominator blows up, and $h(t) \to 0^{-}$.
    \end{itemize}
\end{itemize}
In this indicator design, we introduced an artificial pole at the target SEP, which will be used shortly below as the stability criterion. No other poles are created on the real axis before this artificial one, ensuring clear detection. 

\subsection{Time Contraction Mapping} \label{subsect:3-2}
While the system's governing DAEs~\eqref{eq:DAE} generally do not admit a closed-form solution for $\mathbf{x}(t)$, a sufficiently smooth system can be represented locally by a Taylor series expansion around the initial point $\mathbf{x}(0)$. This power series contains the high-order derivative information we seek. However, its utility is fundamentally restricted by its radius of convergence. As a \textit{local} approximation, it cannot, by itself, determine the \textit{global}, asymptotic behavior of the trajectory as $t \to +\infty$. In practice, even high-order expansions rapidly lose accuracy due to error accumulation and numerical instability, often failing to capture dynamics beyond the first few swings. The conventional strategy of analytic continuation, which involves sequential re-expansions, is precisely the kind of step-by-step process we aim to avoid.

Our approach circumvents this locality problem not by attempting to extend the trajectory in time, but by transforming the temporal domain itself. To this end, we introduce a smooth, monotonically increasing, and bounded function, hereafter termed the \textit{time contraction mapping}
\begin{equation}\label{eq:mapping}
 \tau = \mathcal{M}(t)
\end{equation}
with the following properties:
\begin{enumerate}
    \item It maps the initial time to zero: $\mathcal{M}(0) = 0$.
    \item It maps the infinite time horizon to a finite upper bound $\mathrm{M}$: $\lim_{t \to +\infty} \mathcal{M}(t) = \mathrm{M}$.
    \item $d\mathcal{M}(t) / dt >0$.
\end{enumerate}
This mapping effectively contracts the infinite interval $t \in [0, +\infty)$ into the finite interval $\tau \in [0, \mathrm{M})$.

This transformation is the linchpin of our method. It recasts the intractable problem of analyzing system behavior as $t \to +\infty$ into the well-posed problem of analyzing behavior as $\tau \to \mathrm{M}^{-}$. Consequently, the indicator function from \eqref{eq:h_t} is reformulated in the new time coordinate $\tau$:
\begin{equation}\label{eq:h_tau}
h(\tau) = -\frac{1}{\|\mathbf{x}(\tau) - \mathbf{x}^*\|^2}
\end{equation}

\subsection{Asymptotic Behavior Detection} \label{subsect:3-3}
The stability indicator developed in the previous section now hinges on a single analytical task: detecting whether $h(\tau)$ has a pole at the finite point $\tau = \mathrm{M}$. 
A truncated Taylor series, being a polynomial, is structurally incapable of representing poles. To overcome this limitation, we employ a rational function approximation, which is the natural mathematical structure for modeling poles. Specifically, we use the Padé approximant. 

A Padé approximant of a function $h(\tau)$, denoted $[L/M]$, is a rational function of two polynomials, $P_L(\tau)$ and $Q_M(\tau)$ of degree $L$ and $M$ respectively, whose own Taylor series expansion matches that of $h(\tau)$ up to the highest possible order, $L+M$.
\begin{equation}\label{eq:pade}
[L/M]_{h}(\tau) = \frac{P_L(\tau)}{Q_M(\tau)} = \frac{p_0+p_1\tau+\dots+p_L\tau^L}{q_0+q_1\tau+\dots+q_M\tau^M}
\end{equation}
The coefficients $p_i$ and $q_j$ are determined (up to a scaling factor, typically setting $q_0=1$) such that the Maclaurin series of $[L/M]_{h}(\tau)$ matches that of $h(\tau)$ up to the highest possible order, $L+M$. The critical insight here is that the analytical problem of finding a pole in $h(\tau)$ is transformed into the far more tractable algebraic problem of finding the roots of the denominator polynomial, $Q_M(\tau)$.

The procedure to construct this approximant and detect the pole involves two main steps:
\begin{enumerate}
    \item Compute the Taylor Coefficients of $h(\tau)$:
    The coefficients $\{h_k\}$ required for \eqref{eq:pade} are not computed via direct symbolic differentiation. Instead, we first re-parameterize the original DAEs~\eqref{eq:DAE} in the new time coordinate $\tau$ by applying the chain rule:
    \begin{equation}
    \begin{aligned}
    \frac{\mathrm{d}\mathbf{x}}{\mathrm{d}\tau} &= \theta(\tau)\, \mathbf{f}\big(\mathbf{x}(\tau), \mathbf{v}(\tau)\big) \\
    \mathbf{0} &= \mathbf{g}\big(\mathbf{x}(\tau), \mathbf{v}(\tau) \big)
    \end{aligned}
    \label{eq:transformedDAE_final}
    \end{equation}
    where $\theta(\tau) = dt/d\tau$ is the derivative of the inverse time mapping. We then employ the Differential Transformation (DT) method~\cite{Sun2019power}, a numerical technique that converts these DAEs into a set of algebraic recurrence relations. This allows for the efficient computation of the Taylor coefficients of $\mathbf{x}(\tau)$---and subsequently those of $h(\tau)$---to any desired order.
    
    \item Construct the Approximant and Find the Pole:
    With the coefficients $\{h_k\}$ computed, we construct the $[L/M]$ Padé approximant using \eqref{eq:pade}. We then solve for the roots of the denominator polynomial $Q_M(\tau)$. The smallest positive real root, if one exists, is the estimated location of the dominant pole of $h(\tau)$.
\end{enumerate}

The reliability of this approach is strongly supported by convergence theory. It has been shown that near-diagonal Padé approximants (where $L \approx M$) are particularly effective at capturing the singularity structure of analytic functions \cite{stahl1989convergence, stahl1997convergence}. This provides a solid theoretical foundation for using this technique. In practice, the Padé approximant not only models singularities but also offers a more accurate global representation than a polynomial of the same order, avoiding the oscillatory artifacts (e.g., Runge's phenomenon) that often plague high-order polynomial approximations.

\subsection{Overall Transient Stability Assessment Workflow}

\begin{algorithm}[htbp]
\caption{Transient Stability Assessment Workflow}
\label{alg:Stability}
\textbf{Input:} System model~\eqref{eq:DAE}, time mapping $\mathcal{M}(t)$, initial point $(\mathbf{x}_0, \mathbf{v}_0)$, Padé order $[L/M]$.\\
\textbf{Output:} Stability status (Stable/Unstable).\\
\textbf{Step 1: Initialization}\\
\hspace*{1em} 1.1 Solve for the post-fault SEP $\mathbf{x}^*$ by setting $\dot{\mathbf{x}}=0$ in~\eqref{eq:DAE}.\\
\hspace*{1em} 1.2 Set initial Taylor coefficients: $\mathbf{X}(0) = \mathbf{x}_0$, $\mathbf{V}(0) = \mathbf{v}_0$.

\vspace{0.5em}
\textbf{Step 2: Taylor Coefficient Calculation}\\
\hspace*{1em} \textbf{for} $k = 1$ \textbf{to} $L+M+1$ \textbf{do}\\
\hspace*{3em} Compute the Taylor coefficients $\mathbf{X}(k)$, $\mathbf{V}(k)$ using the DT recurrence relations derived from~\eqref{eq:transformedDAE_final}.\\
\hspace*{1em} \textbf{end for}\\
\hspace*{1em} Compute the Taylor coefficients of the indicator function, $\{h_k\}$ for $k = 0, \dots, L+M$, from the sequence of $\{\mathbf{X}(k)\}$.

\vspace{0.5em}
\textbf{Step 3: Asymptotic Behavior Detection}\\
\hspace*{1em} 3.1 Construct the $[L/M]$ Padé approximant for $h(\tau)$ using the computed coefficients $\{h_k\}$.\\
\hspace*{1em} 3.2 Find all roots of the denominator polynomial $Q_M(\tau)$.\\
\hspace*{1em} 3.3 Identify the smallest positive real root, $\tau_{\text{pole}}$. 

\vspace{0.5em}
\textbf{Step 4: Transient Stability Assessment}\\
\hspace*{1em} \textbf{if} $|\tau_{\text{pole}} - \mathrm{M}| \le \varepsilon$\\
\hspace*{3em} The system is \textbf{Stable}.\\
\hspace*{1em} \textbf{else}\\
\hspace*{3em} The system is \textbf{Unstable}.\\
\hspace*{1em} \textbf{end if}
\end{algorithm}
The complete workflow of the proposed transient stability assessment method is summarized in \textbf{Algorithm~\ref{alg:Stability}}. The process begins with the system's DAE model and initial conditions, transforms the stability problem into the contracted time domain, computes the necessary coefficients for the indicator function, and concludes by checking for a pole at the mapped time horizon using a Padé approximant.

\section{Numerical Study}
To validate the proposed method, we evaluate its performance on three distinct dynamical systems: the Lorenz system, a single machine infinite bus (SMIB) system, and the WSCC 3-machine 9-bus system. For each system, initial conditions are selected from both stable and unstable regions to assess the method's ability to classify system behavior correctly.

The simulation setup is consistent across all studies. The time contraction mapping is defined as $\tau = 1 - (Kt + 1)^{-3}$, which maps $t \in [0, +\infty)$ to $\tau \in [0, 1)$, with the rate parameter set to $K=1$ for the Lorenz system and $K=5$ for the power systems. Since the accuracy requirement grows with the Padé approximant order, the working precision is taken to equal \(L+M+1\) digits. The ground-truth benchmark for comparison is generated using a fourth-order Runge-Kutta (RK4) integrator with a fine time step of $\Delta t = 0.1$\,ms.

The results are presented by reporting the numerically detected pole location, $\tau_{\text{pole}}$, which serves as the direct stability indicator. Additionally, to visually assess the approximation quality, we plot the indicator function $h(\tau)$ as reconstructed by the Padé approximant against the benchmark trajectory. In the subsequent figures, the Padé approximant is denoted in the legend as PA.

\subsection{Lorenz System}
\begin{figure}[tb]
\centering
\begin{subfigure}[d]{0.5\linewidth}
    \includegraphics[width=\linewidth]{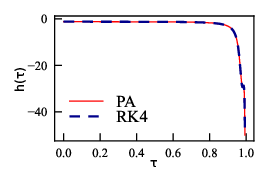}
    \caption{Converge to designated SEP}\label{fig:ls_sep}
\end{subfigure}\hfill
\begin{subfigure}[d]{0.5\linewidth}
    \includegraphics[width=\linewidth]{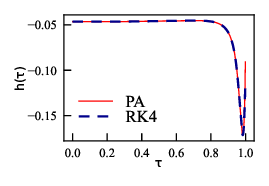}
    \caption{Converge to other SEP}\label{fig:ls_other}
\end{subfigure}\hfill
\caption{Approximant performance of stable Lorenz}
\label{fig:lorenz}
\end{figure}

\setlength{\abovecaptionskip}{0pt} 
\setlength{\textfloatsep}{0pt} 
\begin{figure}[tb]
\centerline{\includegraphics{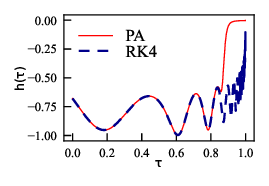}}
\caption{Approximant performance of chaotic Lorenz}
\label{fig:ls_chaos}
\end{figure}
First, we configure the system with stable parameters ($\sigma = 1, \rho = 2, \beta = 1$) and test two initial conditions. As shown in Fig.~\ref{fig:lorenz}, a $[40/40]$ Padé approximant is used, and its reconstructions show excellent agreement with the benchmarks. For the case converging to the designated SEP (Fig.~\ref{fig:ls_sep}), the method correctly identifies stability by detecting a pole at $\tau_{\text{pole}} = 0.996$ (0.4\% error from the theoretical location at M=1). For the case converging to another SEP (Fig.~\ref{fig:ls_other}), the approximant correctly settles at a finite value, confirming a non-divergent but off-target trajectory.

Next, the parameters are set to their classic chaotic values ($\sigma = 10, \rho = 28, \beta = 8/3$). As shown in Fig.~\ref{fig:ls_chaos}, the indicator function $h(\tau)$ for this chaotic trajectory exhibits sustained, non-periodic oscillations. The Padé approximant accurately captures the initial cycles of this complex behavior. Constrained by its finite order, the approximant cannot replicate the bounded, non-convergent nature of a chaotic attractor indefinitely. Instead, it correctly interprets the trajectory as unstable with respect to the designated SEP, with its value tending towards zero, the expected behavior for a non-convergent case.

\subsection{SMIB System}
\begin{figure}[tb]
\centering
\begin{minipage}[t]{0.5\linewidth} 
    \centering
    \includegraphics[scale=0.8]{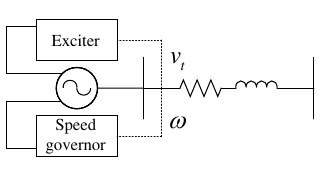}
    \caption{SMIB system}
    \label{fig:SMIBTopology}
\end{minipage} \hfill
\begin{minipage}[t]{0.45\linewidth} 
    \centering
    \includegraphics[scale=0.8]{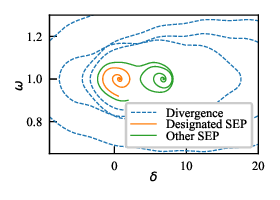}
    \caption{$\delta-\omega$ phase plane}
    \label{fig:SMIB_behavior}
\end{minipage}
\end{figure}

\begin{figure}[tb]
\centering
\begin{subfigure}[d]{0.5\linewidth}
    \includegraphics[width=\linewidth]{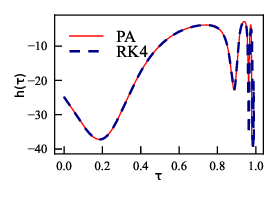}
    \caption{Converge to designated SEP}\label{fig:SMIB_sep}
\end{subfigure}\hfill
\begin{subfigure}[d]{0.5\linewidth}
    \includegraphics[width=\linewidth]{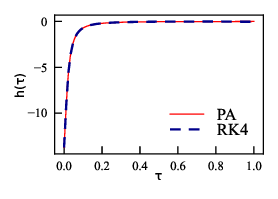}
    \caption{Converge to other SEP}\label{fig:SMIB_other}
\end{subfigure}\hfill
\begin{subfigure}[d]{0.5\linewidth}
    \includegraphics[width=\linewidth]{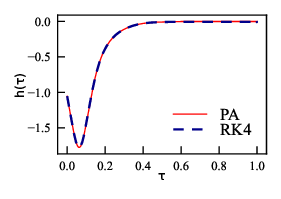}
    \caption{Diverge}\label{fig:SMIB_div}
\end{subfigure}\hfill
\caption{Approximant performance for SMIB}
\label{fig:SMIB}
\end{figure}
The SMIB system adopted in this section is shown in Fig.~\ref{fig:SMIBTopology}. 
The system, depicted in Fig.~\ref{fig:SMIBTopology}, consists of a synchronous generator connected to an infinite bus via a transmission line. The generator is represented by a sixth-order model, equipped with an IEEE Type-1 exciter and a simplified linear speed governor~\cite{demetriou2015dynamic}. To ensure the smoothness of the governing equations as required by our method, the operational limits of the controllers are neglected in this study.

The dynamic behavior of the SMIB system under different initial conditions is illustrated by the phase plane portrait in Fig.~\ref{fig:SMIB_behavior}. We select initial conditions for three key scenarios: convergence to the designated SEP, convergence to another SEP, and divergence. The results, obtained using a $[100/100]$ Padé approximant, are presented in Fig.~\ref{fig:SMIB}. The method demonstrates high fidelity, correctly capturing the distinct asymptotic behavior for all three cases.

\subsection{WSCC 3-machine 9-bus System}
\begin{figure}[tb]
\centering
\includegraphics[scale=0.8]{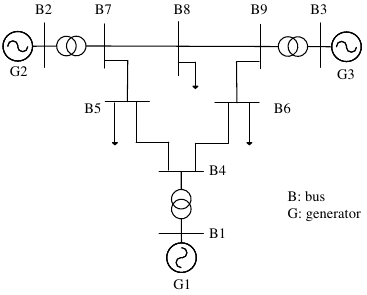}
\caption{WSCC 3-machine 9-bus System}
\label{fig:3m9b}
\end{figure}

\begin{figure}[tb]
\centering
\begin{subfigure}[d]{0.5\linewidth}
    \includegraphics[width=\linewidth]{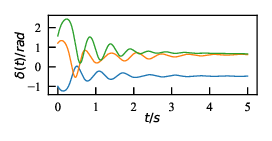}
    \caption{Cleared at 0.25~s}\label{fig:time1.25}
\end{subfigure}\hfill
\begin{subfigure}[d]{0.5\linewidth}
    \includegraphics[width=\linewidth]{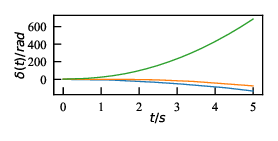}
    \caption{Cleared at 0.26~s}\label{fig:time1.26}
\end{subfigure}\hfill
\caption{Swing curves for different FCTs}\label{fig:3m9b_Time}
\end{figure}

\begin{figure}[tb]
\centering
\begin{subfigure}[d]{0.5\linewidth}
    \includegraphics[width=\linewidth]{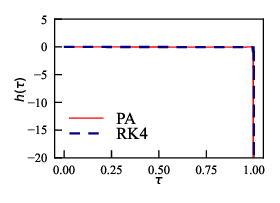}
    \caption{FCT = 0.25\,s}\label{fig:3m9b_1.25}
\end{subfigure}\hfill
\begin{subfigure}[d]{0.5\linewidth}
    \includegraphics[width=\linewidth]{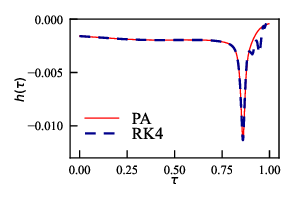}
    \caption{FCT = 0.26\,s}\label{fig:3m9b_1.26}
\end{subfigure}\hfill
\caption{Approximant performance for 9-bus system}
\label{fig:3m9b_1.25+1.26}
\end{figure}
The system topology is shown in Fig.~\ref{fig:3m9b}. The test scenario involves a three-phase fault applied at Bus 9 through an impedance of $R_f + jX_f = 0.01 + j0.02$ p.u.

To investigate the method's performance near the stability boundary, we analyze the system's response to different fault-clearing times (FCTs). Time domain simulations, shown in Fig.~\ref{fig:3m9b_Time}, reveal that the critical clearing time for this fault is between 0.25\,s and 0.26\,s. The system remains stable if the fault is cleared at 0.25\,s but loses synchronism if the FCT is extended to 0.26\,s.

The $[100/100]$ Padé approximant correctly predicts the outcome for both the stable (when FCT = 0.25\,s) and unstable (when FCT = 0.26\,s) scenarios, as shown in Fig.~\ref{fig:3m9b_1.25+1.26}. For the unstable case in particular, a visible deviation from the benchmark appears at the final stage. This is an expected artifact of a finite-order model capturing a divergent trajectory, but the essential instability verdict is correctly identified.



\section{Discussions}
\textbf{Error tolerance in the asymptotic sense:} The finite truncation of the Taylor series and the associated Padé approximant will eventually lead to a gap between the indicator function and its approximation. However, this discrepancy is tolerable even if it grows to infinity, as long as the associated pole exists, as demonstrated by the simulation results above. 

\textbf{Parallel processing opportunity:} Conceptually, all the higher-order derivatives of the differential equation system can be evaluated in parallel, since they are all computed at the same point without any temporal dependence. 

\textbf{Promise in general applications:} This approach not only holds significant promise for power system stability assessment but also can be extended to the stability analysis of a broad class of nonlinear dynamical systems.

\bibliography{ref}

@article{chiang2002study,
  title={Study of the existence of energy functions for power systems with losses},
  author={Chiang, H-D},
  journal={IEEE Transactions on Circuits and Systems},
  volume={36},
  number={11},
  pages={1423--1429},
  year={2002},
  publisher={IEEE}
}

@article{wu2019holomorphic,
  title={Holomorphic embedding based continuation method for identifying multiple power flow solutions},
  author={Wu, Dan and Wang, Bin},
  journal={IEEE access},
  volume={7},
  pages={86843--86853},
  year={2019},
  publisher={IEEE}
}

@article{kundur2007power,
  title={Power system stability},
  author={Kundur, Prabha and others},
  journal={Power system stability and control},
  volume={10},
  number={1},
  pages={7--1},
  year={2007}
}

@article{demetriou2015dynamic,
  title={Dynamic IEEE test systems for transient analysis},
  author={Demetriou, Panayiotis and Asprou, Markos and Quiros-Tortos, Jairo and Kyriakides, Elias},
  journal={IEEE Systems Journal},
  volume={11},
  number={4},
  pages={2108--2117},
  year={2015},
  publisher={IEEE}
}

@misc{hairer1996solvingODE,
  title={Solving ordinary differential equations. II, volume 14 of Springer Series in Computational Mathematics},
  author={Hairer, Ernst and Wanner, Gerhard},
  year={1996},
  publisher={Springer-Verlag, Berlin,}
}

@article{Sun2019power,
  title={Power system time domain simulation using a differential transformation method},
  author={Liu, Yang and Sun, Kai and Yao, Rui and Wang, Bin},
  journal={IEEE Transactions on Power Systems},
  volume={34},
  number={5},
  pages={3739--3748},
  year={2019},
  publisher={IEEE}
}

@article{bellizio2022changingtopology,
  title={Machine-learned security assessment for changing system topologies},
  author={Bellizio, Federica and Cremer, Jochen L and Strbac, Goran},
  journal={International Journal of Electrical Power \& Energy Systems},
  volume={134},
  pages={107380},
  year={2022},
  publisher={Elsevier}
}

@article{guidotti2018survey,
  title={A survey of methods for explaining black box models},
  author={Guidotti, Riccardo and Monreale, Anna and Ruggieri, Salvatore and Turini, Franco and Giannotti, Fosca and Pedreschi, Dino},
  journal={ACM computing surveys (CSUR)},
  volume={51},
  number={5},
  pages={1--42},
  year={2018},
  publisher={ACM New York, NY, USA}
}

@article{li2022deep,
  title={A deep-learning intelligent system incorporating data augmentation for short-term voltage stability assessment of power systems},
  author={Li, Yang and Zhang, Meng and Chen, Chen},
  journal={Applied Energy},
  volume={308},
  pages={118347},
  year={2022},
  publisher={Elsevier}
}

@article{sarajcev2022artificial,
  title={Artificial intelligence techniques for power system transient stability assessment},
  author={Sarajcev, Petar and Kunac, Antonijo and Petrovic, Goran and Despalatovic, Marin},
  journal={Energies},
  volume={15},
  number={2},
  pages={507},
  year={2022},
  publisher={MDPI}
}

@article{stahl1989convergence,
  title={On the convergence of generalized Pad{\'e} approximants},
  author={Stahl, Herbert},
  journal={Constructive Approximation},
  volume={5},
  number={1},
  pages={221--240},
  year={1989},
  publisher={Springer}
}

@article{stahl1997convergence,
  title={The convergence of Pad{\'e} approximants to functions with branch points},
  author={Stahl, Herbert},
  journal={Journal of Approximation Theory},
  volume={91},
  number={2},
  pages={139--204},
  year={1997},
  publisher={Elsevier}
}

@article{magnusson1947transient,
  title={The transient-energy method of calculating stability},
  author={Magnusson, Philip Cooper},
  journal={Transactions of the American Institute of Electrical Engineers},
  volume={66},
  number={1},
  pages={747--755},
  year={1947},
  publisher={IEEE}
}

@book{chiang2011direct,
  title={Direct methods for stability analysis of electric power systems: theoretical foundation, BCU methodologies, and applications},
  author={Chiang, Hsiao-Dong},
  year={2011},
  publisher={John Wiley \& Sons}
}

@article{caliskan2014compositional,
  title={Compositional transient stability analysis of multimachine power networks},
  author={Caliskan, Sina Yamac and Tabuada, Paulo},
  journal={IEEE Transactions on Control of Network systems},
  volume={1},
  number={1},
  pages={4--14},
  year={2014},
  publisher={IEEE}
}

@article{dunstan1954digital,
  title={Digital load flow studies [includes discussion]},
  author={Dunstan, Lyle A},
  journal={Transactions of the American Institute of Electrical Engineers.},
  volume={73},
  number={1},
  pages={825--832},
  year={1954},
  publisher={IEEE}
}

@article{chang1995numericaldirect,
  title={Direct stability analysis of electric power systems using energy functions: theory, applications, and perspective},
  author={Chang, Hsiao-Dong and Chu, Chia-Chi and Cauley, Gerry},
  journal={Proceedings of the IEEE},
  volume={83},
  number={11},
  pages={1497--1529},
  year={1995},
  publisher={IEEE}
}

@misc{dan2019solutions,
  author       = {D. Wu and B. Wang},
  title        = {Collection of numerous power flow solutions of standard {IEEE} test systems},
  howpublished = {IEEE Dataport},
  year         = {2019},
  doi          = {10.21227/24bh-hj72},
  note         = {\emph{Available: https://dx.doi.org/10.21227/24bh-hj72}}
}

\end{document}